\documentstyle[editedbook,epsfig,psfig,epsf]{mq}

\def\cm2{cm$^2$ }
\def\se1{s$^{-1}$ }

\def\la{\mathrel{\mathchoice {\vcenter{\offinterlineskip\halign{\hfil
$\displaystyle##$\hfil\cr<\cr\sim\cr}}}
{\vcenter{\offinterlineskip\halign{\hfil$\textstyle##$\hfil\cr
<\cr\sim\cr}}}
{\vcenter{\offinterlineskip\halign{\hfil$\scriptstyle##$\hfil\cr
<\cr\sim\cr}}}
{\vcenter{\offinterlineskip\halign{\hfil$\scriptscriptstyle##$\hfil\cr
<\cr\sim\cr}}}}}

\begin{opening}
\title{Assessing the X-ray Contribution from Jets in X-ray Binaries}
\vspace{-2cm}\author{S. Markoff$^{1,*}$, M. Nowak$^{2}$, S. Corbel$^3$,
R. Fender$^4$ \& H. Falcke$^1$}
\institute{$^1$ Max-Planck Institut f\"ur Radioastronomie, Auf dem H\"ugel
69, D-53121 Bonn, Germany
\\$^2$ Massachusetts Institute of Technology, Center for Space Research
Rm. NE80-6077,
77 Massachusetts Ave., Cambridge, MA 02139, USA
\\$^3$ Universit\'e Paris VII and Service d'Astrophysique, CEA,
CE-Saclay, 91191 Gif sur Yvette, France
\\$^4$ Astronomical Institute `Anton Pannekoek' and Center for High
Energy Astrophysics, University of Amsterdam, 
Kruislaan 403, 1098 SJ Amsterdam, The Netherlands
\\$^*$ Humboldt Research Fellow}
\end{opening}

\runningtitle{X-ray Contribution of XRB Jets}
\runningauthor{Markoff et al.}

\begin{document}
\vspace{-0.5cm}
\begin{abstract}
{\small Astrophysical jets exist from the stellar scale up to AGN, and seem to
share common features particularly in the radio.  But while AGN jets
are known to emit X-rays, the situation for XRB jets is not so clear.
Radio jets have been resolved in several XRBs in the low/hard state,
and it seems likely that some form of outflow is present whenever this
state is achieved.  Interestingly, the flat-to-inverted radio
synchrotron emission associated with these outflows strongly
correlates with the X-ray emission in several sources, suggesting that
the jet plasma plays a role at higher frequencies.  In this same
state, there is also increasing evidence for a turnover in the
IR/optical where the flat-to-inverted spectrum seems to connect to an
optically thin component extending into the X-rays.  We discuss how
jet synchrotron emission is likely to contribute to the X-rays, in
addition to inverse Compton up-scattering, providing a natural
explanation for these correlations and the turnover in the IR/optical
band.}
\end{abstract}

\section{Introduction}

Active galactic nuclei (AGN) jets have been extensively imaged in the
radio, and also emit significantly at higher frequencies including the
X-rays via synchrotron and inverse Compton (IC).  It turns out that
black hole candidate (BHC) X-ray binaries (XRBs) also produce
collimated outflows, at least when in the low/hard state (LHS)
\cite{Fender2001a}.  The jets in several persistent Galactic sources
have already been resolved in the radio (e.g., 1E1740.7-2942,
\cite{Mirabeletal1992}; Cyg X-1, \cite{Stirlingetal2001}).

Analogous to the ``signature'' emission of
 compact radio cores in AGN (e.g.,
\cite{BlandfordKoenigl1979}), XRB jets contribute a flat-to-inverted
radio synchrotron component to the LHS spectra.  Beyond this radio
signature, a typical LHS spectrum shows a weak thermal
contribution and a hard power-law at higher frequencies.  This
has generally been interpreted in terms of a Standard Thin Disk (SD;
\cite{ShakuraSunyaev}) which either transitions at some radius to an
optically thin, non-radiative flow, or is underlying a
corona (for reviews see \cite{Poutanen,NowakWilmsDove}).  The hotter
plasma is believed to account for the hard power-law via IC
upscattering of the thermal SD photons.  

While variations of this picture can successfully explain the X-ray
features, there is increasing evidence that---at least in the LHS and
likely also the quiescent state---the jet is also playing a role
outside the radio band.  The extent of this role is not yet clear, but
we have found that emission from the jets alone can actually account
for the majority of the broadband LHS spectrum (excluding the thermal
SD contribution) in sources where simultaneous radio, X-ray and
sometimes infrared (IR)/optical data are available.  The models for
these various sources also show a surprising degree of similarity in
their input parameters, and is the only model yet which can explain
the multi-wavelength correlations which are now being seen in several
sources.  It is therefore important to begin exploring ways in which
the jets can be incorporated into the previously disk/corona-only X-ray
picture, and to know the extent to which they can reasonably
contribute.  

\section{Brief Model Summary}

We first considered a jet model for the BHC XTE~J1118+480, because of
the excellent broadband observational coverage.  Details of the model,
data references and figures can be found in
\cite{MarkoffFalckeFender}.  The basic idea is that a small fraction
of the charge-neutral accreting plasma from the inner disk, which we
take to be optically thin and hot, escapes the black hole and is
collimated into a jet.  The energy in the jet is divided equally
between the kinetic energy and the internal energy, dominated by hot
electrons and the magnetic fields.  Under these assumptions, to
explain the radio/IR spectra with a realistic jet power, $Q_{\rm
j}=q_{\rm j}\dot M c^2$ where $q_{\rm j}\sim0.01-0.1$, the thermal
plasma must encounter some acceleration region within the jet (see,
e.g., \cite{Markoffetal2002}).  This process is also inferred because
of the observed optically thin synchrotron power-law seen during radio
outbursts, indicating the presence of nonthermal particles.

The question of whether jet synchrotron will contribute to the
X-rays is thus a matter of the maximum energy to which the particles
can be accelerated.  Synchrotron cooling will dominate as long as the
SD does not extend too close to the base of the jet ($\la 10 r_g$).
In the specific case of Fermi acceleration, the location of the
maximum cutoff is not dependent on the magnetic field, the jet power,
or the shock location but is instead dependent on two plasma
parameters: the mean free path for diffusive scattering and the speed
of the shock in the plasma frame.  Because we would expect XRBs to
have similar shock structures, we should get roughly similar cutoffs
for different sources and accretion rates.  In cases where the photon
field from the SD is high enough, however, IC from the jet could
instead dominate.

\section{Some Applications}
 
\begin{figure}[t]
\centerline{\psfig{figure=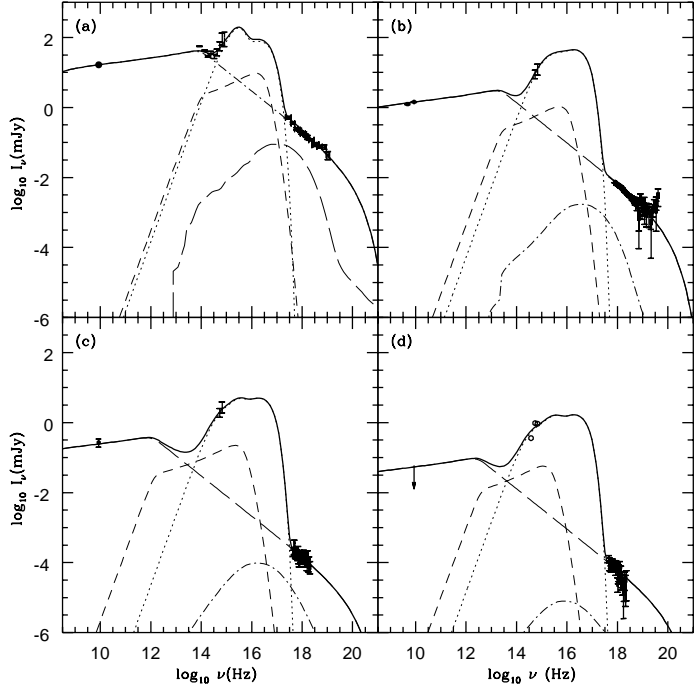,width=10cm}}
\caption[]{Jet model fit to four representative broadband spectra for
GX~339-4, with details and data references in \cite{Markoffetal2002}.
Solid thick line: total spectrum, dotted line: multi-temperature
blackbody outer disk plus single blackbody irradiation contribution,
short-dashed: synchrotron emission from the jet before acceleration,
dot-dashed: synchrotron emission from jet regions after acceleration,
and long-dashed: inverse Compton upscattered disk and jet photons.
(a) 1981 IR/X-ray data set, with extrapolated radio point, (b)
May 14, 1999 data set, (c) August 17, 1999 data set, (d) Sept. 15,
2000 data set (in quiescence), radio is a 3$\sigma$ upper limit.}\label{fig1}  
\end{figure}

We have applied this model so far to 4 LHS sources with simultaneous
or quasi-simultaneous data, in addition to XTE~J1118+480.  One of the
most interesting data sets comes from the 1981 observations of the
Galactic BHC GX~339-4, presented in \cite{CorbelFender2002}.  The
IR/optical wavebands seem to indicate the clear presence of both jet
and disk components, which has not been seen explicitly in any other
source.  The first component at lower frequencies has a negative
slope, suggesting that we are seeing the expected turnover from the
optically thick to optically thin regimes.  The shape of this spectrum
at higher frequencies is hidden under a component which is likely due
to thermal emission from the SD, as indicated by the sharp rise in the
optical points.  However, if the simultaneously measured X-rays are
traced back to the IR, they line up with the turnover remarkably well,
supporting their interpretation as synchrotron emission.  We present
one possible model fit in Fig.~\ref{fig1}a.

A special feature of GX~339-4 is that its radio emission has been
observed to tightly correlate with the X-ray emission over the entire
range of LHS luminosities, and also down to a very weak level of
emission in the ``off'' state \cite{Corbeletal2000,Corbeletal2002}.
This also suggests that the plasma emitting the synchrotron
radiation plays a role at higher energy.  In Fig.~\ref{fig1}b-d we
show representative fits to three of the 13 simultaneous data sets (see
\cite{Markoffetal2002}), which cover almost three orders of
magnitude in X-ray luminosity.  

\begin{figure}[t]
\centerline{\psfig{figure=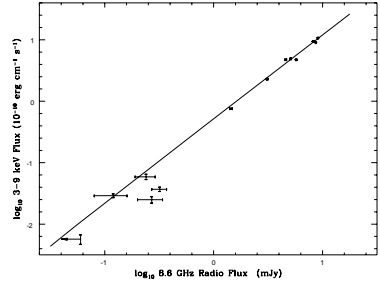,width=4cm,angle=-90}}
\caption[]{The jet model-predicted radio (8.6 GHz)/X-ray (integrated
3-9 keV emission) correlation (solid line).  The data are from
\cite{Corbeletal2000,Corbeletal2002}.  The only changing parameter is
the jet power.  The arrow represents 3$\sigma$ upper limits.  See
\cite{Markoffetal2002} for details.}\label{fig2}
\end{figure}

The correlation data of \cite{Corbeletal2000,Corbeletal2002} can be
seen explicitly in Fig.~\ref{fig2}.  The solid line is the prediction
of the synchrotron-dominated jet model, with the power as the only
changing parameter.  The two ``outliers'' may be due to a reflaring at
the time of the observation.  If we ignore them, the slope of the
correlation is $\sim1.4$, which is predicted analytically from the
dependence of the radio and X-ray luminosities on jet power in our
model.

\begin{figure}[h]
\centerline{\hbox{\psfig{figure=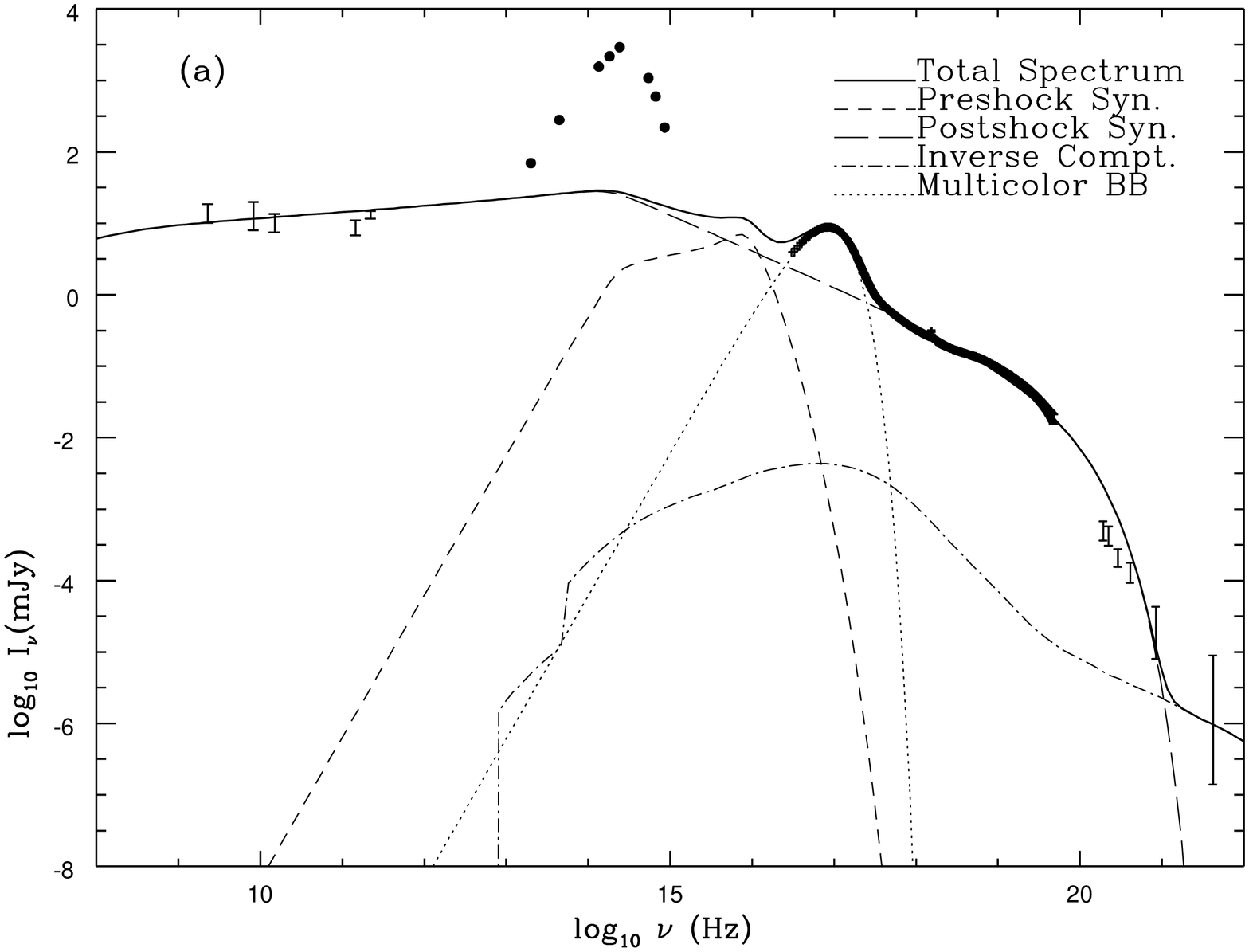,width=.37\textwidth,angle=-90}\hfill\psfig{figure=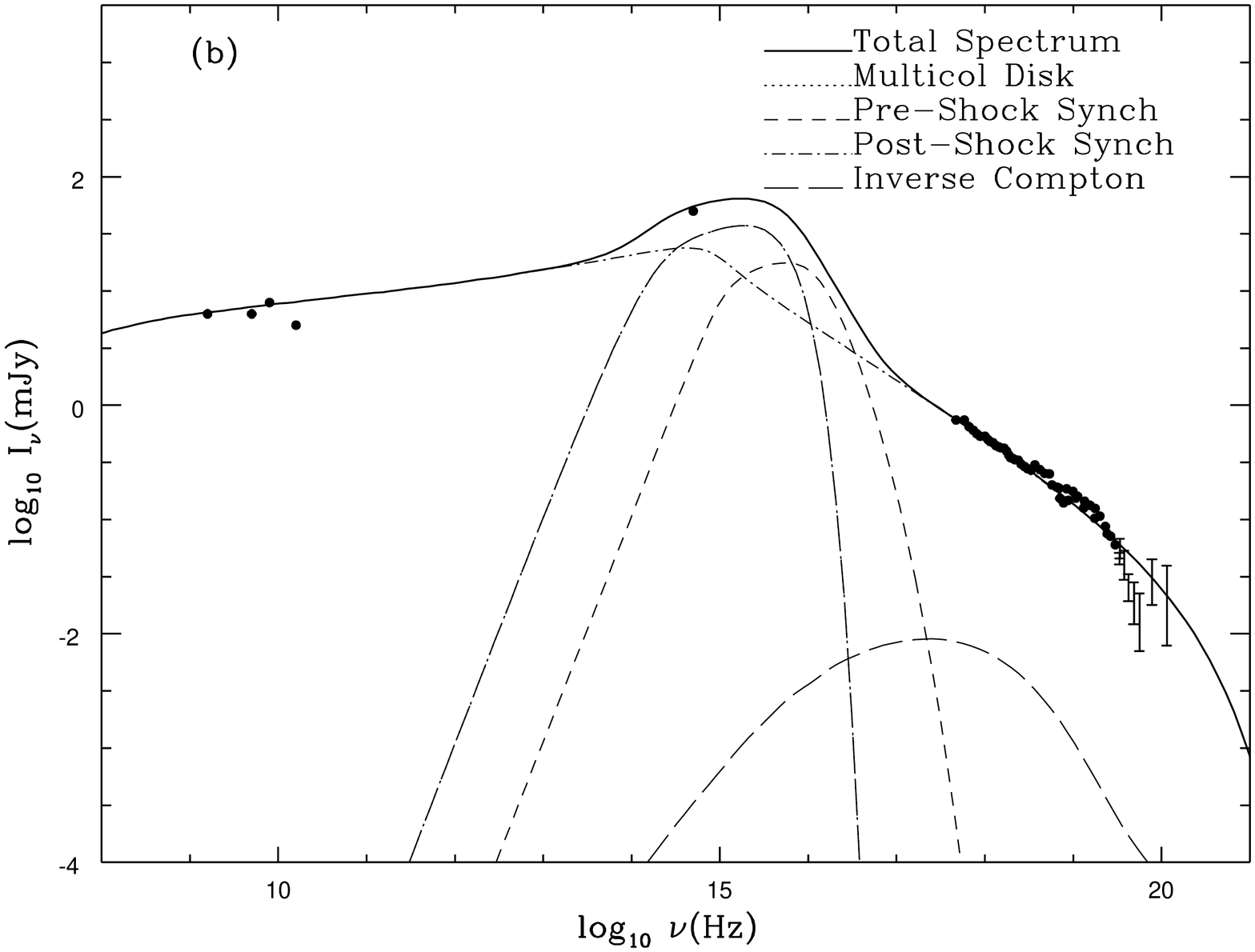,width=.37\textwidth,angle=-90}}}
\caption[]{Two other applications: (a) Cyg X-1, non-simultaneous LHS
data compiled by S. Tigelaar, the IR excess is due to the companion;
(b) GRO~J0422+32, simultaneous radio
and IR compiled by C. Brocksopp, X-rays quasi-simultaneous from
Mir-Kvant but averaged over 5 decaying days by \cite{Sunyaev}.}
\label{fig3}
\end{figure}

In Fig.~\ref{fig3} we show two other applications which are not as
well constrained due to the non-simultaneity.  Simultaneous radio and
X-ray data are critical for understanding this model.  At this
preliminary stage, however, we seem to be discovering a few trends
beyond the correlations.  For instance, in all sources the shock
acceleration zone seems to fall $\sim 10^2$ $r_g$, which is roughly
where the reconfinement shock is thought to occur in AGN (e.g.,
\cite{Birettaetal}).  This means that the ``turnover coincidence'' where
the X-rays always trace back to the IR could be a signature of
acceleration from this shock zone.

\end{document}